# A Novel Approach for Image Steganography in Spatial Domain

By Fatema Akhter

Jatiya Kabi Kazi Nazrul Islam University, Bangladesh

*Abstract -* This paper presents a new approach for hiding information in digital image in spatial domain. In this approach three bits of message is embedded in a pixel using Lucas number system but only one bit plane is allowed for alternation. The experimental results show that the proposed method has the larger capacity of embedding data, high peak signal to noise ratio compared to existing methods and is hardly detectable for steganolysis algorithm.

*GJCST-F Classification:* I.4.0

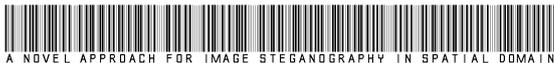

Strictly as per the compliance and regulations of:

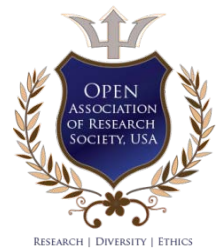

# A Novel Approach for Image Steganography in Spatial Domain

Fatema Akhter

*Abstract -* This paper presents a new approach for hiding information in digital image in spatial domain. In this approach three bits of message is embedded in a pixel using Lucas number system but only one bit plane is allowed for alternation. The experimental results show that the proposed method has the larger capacity of embedding data, high peak signal to noise ratio compared to existing methods and is hardly detectable for steganolysis algorithm.

## I. Introduction

In recent year Steganography is considered as a promising way of safe electronic communication. Steganography is the art and science of hiding information such its presence cannot be detected and a communication is happening. The word steganography comes from the Greek words Steganos (covered) and graphia (writing) [1] and literally means covered writing[2]. Using steganography, information is embedded in a medium such as image, audio, video or text file called carrier in a way that it is not detectable by others [3].

There are two popular techniques regularly used for information hiding, the spatial domain and frequency domain. In spatial domain the information bit is inserted directly and embedded in the intensity of the cover image pixel while in frequency domain the cover image is first transformed to frequency domain and information is hidden in wavelet [4]. This paper focuses on digital image and spatial domain. When designing a steganogrphic algorithm, three properties are to be carefully considered, embedding capacity (payload), image distortion and undetectability. A proper trade off needed as larger information is embedded in cover image more detectable artifacts would be introduced in stegos and would cause more image distortion.

This paper proposes a new approach in spatial domain using Lucas number represenation where three bits of message is embedded without degrading the quality of image. Decomposition of image pixel using Lucas number allows higher bit plane for embedding information that increases the embedding capacity. Although three bits will be embedded in each pixel, only one bit plane have to be changed. The embedded bits can take any place among seven least significant bits which makes it hardly detectable for existing steganolysis algorithms. The rest of the paper is organized as follows: Section 2 presents some related works and algorithms of steganography. Section 3 introduces our proposed stenographic method. Section 4 describes the experimental results and presents comparisons with other works. Finally this paper is concluded with Section 5.

## II. Related Works

This section gives a brief overview of the typical spatial domain algorithms including LSB replacement [5], LSB matching [6], LSBM revisited [7] approaches. LSB replacement algorithm takes the advantage of human eye [8]. An image is consists of many pixels and each pixel can be expressed by a number between 0 to 255, in fact can be represented by 8 bits. It is been noticed that the LSB of a pixel contains no visual information and human eye never realizes a change in this bit. The information bit is simply overwrites the LSB i.e. the first bit plane. LSBM is very similar to LSB replacement algorithm. For LSBM, if the secret bit is not equal to the LSB of the given pixel then ±1 is added randomly to the pixel while keeping the altered pixel value in the range of [0,255]. Using Pseudo Random Number Generator secret bits are scattered on the cover image.

In LSB only the LSB bit plane is altered as a result a structural asymmetry is easily visible to steganolysis algorithms [9] like Chi-Squared attack [10], Chen[11] and Regular/Singular(RS) groups analysis[12] and Sample pair analysis[13]. In LSBM the probability of increasing and decreasing for each modified pixel is same and so the asymmetry artifacts introduced by LSB replacement is avoided. So common approaches used to detect LSB replacement are totally ineffective at detecting the LSBM. But with time several steganolysis algorithms [14-16] have been proposed to analyze the LSBM.

Unlike LSB and LSBM, LSB matching revisited(LSBMR) uses a pair of pixels as an embedding unit in which the LSB of the first pixel carries one bit message and the relationship (odd - even combination) of the two consecutive pixels carries another bit of message. The modification rate of pixels is decreased from 0.5 to 0.375 bits/pixel. This can avoid the LSB replacement asymmetry and make the detection slightly more difficult than the LSBM approaches.

*Author :* B.Sc Engineering in the Department of Computer Science and Engineering at Jatiya Kabi Kazi Nazrul Islam University, Bangladesh.
*E-mail :* fatema_kumu@yahoo.com

Based on the existing works and experimental results, the weakness of LSB, LSBM and LSBR is analyzed here and this paper proposes a new approach that is easy to implement, avoids asymmetry of LSB replacement, increases the embedding capacity 3 bits per pixel (bpp), decreases the modification rate to 1/8 = 0.125 and most of well-known steganolysis attacks are unable to detect the existence of secret message embedded using our algorithm. Figure 1 shows cover image and stego image using our method at 30% embedding rate.

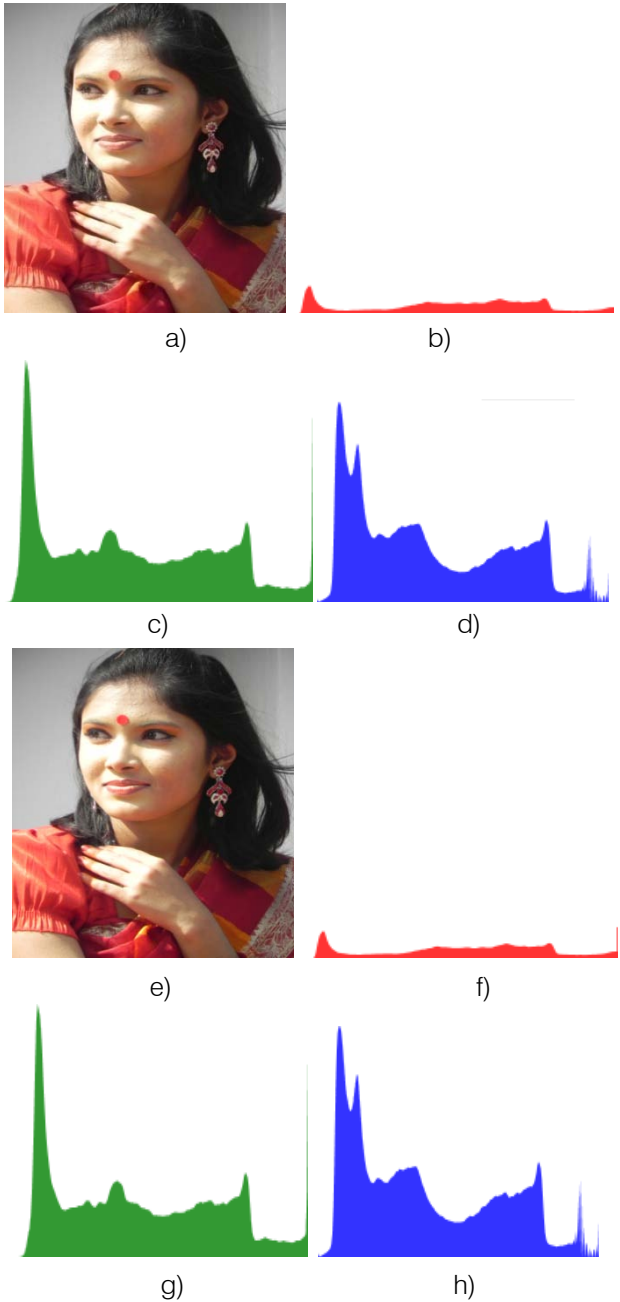

Figure 1 : Cover image and stego image with histogram for finding visual artifact introducing our method, a) Cover Image, b) Histogram of cover image red, c) Green, d) Blue, e) Stego image, f) Histogram of stego image red, f) Green and g) Blue

## III. Proposed Approach

This section describes our proposed approach that embeds 3 bits in a pixel but alters only one bit plane. First the image pixel is decomposed using Lucas number [17] presentation. The Lucas number sequence is generated using the following formula

$$L_n = L_{n-1} + L_{n-2} \text{ where } L_1 = 2 \text{ and } L_2 = 1$$

To represent a number in the range [0,255] we need twelve Lucas numbers.

| L12 | L11 | L10 | L9 | L8 | L7 | L6 | L5 | L4 | L3 | L2 | L1 |
|-----|-----|-----|----|----|----|----|----|----|----|----|----|
| 199 | 123 | 76  | 47 | 29 | 18 | 11 | 7  | 4  | 3  | 1  | 2  |

The image pixel value will be represented as sum of non consecutive Lucas numbers. For example binary and Lucas presentation of pixel value 26 is given below.

Binary representation 00011010
Lucas number representation 000001010010

The motivation behind using Lucas number is that Lucas bit is less significant than that of traditional binary bit. For example if the 5th bit of pixel value 26 is altered to 0 then it becomes 10 in binary representation while 19 in Lucas representation which is more close to the original value.

Binary representation 00001010
Lucas number representation 000001000010

A 7 bit encoder/decoder is proposed for generating stego image. It takes 7 bits ($b_7, b_6, b_5, b_4, b_3, b_2, b_1$) as inputs and produces 3 bits of outputs $f_3, f_2,$ and $f_1$. Figure 2 shows the encoder/decoder generator.

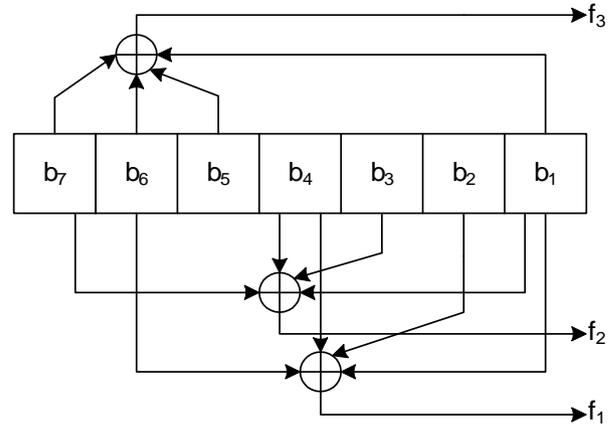

Figure 2 : Encoder/Decoder for stego image generation

Table 1 : The procedure of changing seven LSB of a pixel for three bits of message

| Seven LSB bits | Encoder/Decoder Output | Changes in cover image to match desired message | |
|---|---|---|---|
| | | Message (3 bits) | Stego Pixel (7bits) |
| 0000000 | 0 0 0 | 0 0 1 | 0 0 0 0 0 1 0 |
| | | 0 1 0 | 0 0 0 0 1 0 0 |
| | | 0 1 1 | 0 0 0 1 0 0 0 |
| | | 1 0 0 | 0 0 1 0 0 0 0 |
| | | 1 0 1 | 0 1 0 0 0 0 0 |
| | | 1 1 0 | 1 0 0 0 0 0 0 |
| | | 1 1 1 | 0 0 0 0 0 0 1 |
| 0000001 | 1 1 1 | 0 0 0 | 0 0 0 0 0 0 0 |
| | | 0 0 1 | 1 0 0 0 0 0 1 |
| | | 0 1 0 | 0 1 0 0 0 0 1 |
| | | 0 1 1 | 0 0 1 0 0 0 1 |
| | | 1 0 0 | 0 0 0 1 0 0 1 |
| | | 1 0 1 | 0 0 0 0 1 0 1 |
| | | 1 1 0 | 0 0 0 0 0 1 1 |
| - - - - - - - | - - - | - - - | - - - - - - - |
| 1111111 | 0 0 0 | 0 0 1 | 1 1 1 1 1 0 1 |
| | | 0 1 0 | 1 1 1 1 0 1 1 |
| | | 0 1 1 | 1 1 1 0 1 1 1 |
| | | 1 0 0 | 1 1 0 1 1 1 1 |
| | | 1 0 1 | 1 0 1 1 1 1 1 |
| | | 1 1 0 | 0 1 1 1 1 1 1 |
| | | 1 1 1 | 1 1 1 1 1 1 0 |

For each pixel the 7 least significant bits are applied to the encoder/decoder, three Exclusive OR operations create three outputs $f_3$, $f_2$ and $f_1$ where

$$f_3 = b1 \oplus b5 \oplus b6 \oplus b7$$
$$f_2 = b1 \oplus b3 \oplus b4 \oplus b7$$
$$f_1 = b1 \oplus b2 \oplus b4 \oplus b6$$

Suppose $f_3$, $f_2$, $f_1$ are the hidden message which is embedded in the pixel. If $f_3$, $f_2$, $f_1$ are the same as the hidden message then there is no need to alter the original image. If not, it is needed to change the original image in a way to cause the output of the encoder/decoder be equal to the hidden message. This paper allows alternation in only one bit plane that can be any of the seven bits to make the outputs of the encoder/decoder is equal to the hidden message. The procedure that changes only one bit is described in table I.

IV. Experimental Results

In this section some experimental results are presented to demonstrate the effectiveness of the proposed method compared with existing relevant methods. USC–SIPI [18] and UCID [19] image datasets are used for the performance evaluation.

a) Embedding Capacity

As explained in the previous section, 3 bits of message can be embedded in this method. So the capacity of embedded data is increased up to 3 bits/pixel which is thrice the capacity of existing LSB replacement, LSB matching and LSBM revisited approaches. So our method outperforms all the existing algorithms in embedding capacity and payload.

b) Image Quality Analysis

The quality of the image is evaluated by Peak Signal to Noise Ratio (PSNR) [20] where it is defined as:

$$PSNR = 10 log_{10} \frac{I^2_{MAX}}{MSE}$$

Where $I^2_{MAX}$ is equal to 255 as the maximum possible value of a pixel in the gray scale images or represents the maximum possible value of a color in the color images. Mean Square Error (MSE) for the gray scale images is computed as follows

$$MSE = \frac{1}{R*C} \sum_{i=1}^{R} \sum_{j=1}^{C} (|I_{i,j} - I'_{i,j}|)^2$$

For color image,

$$MSE = \frac{1}{R*C*L} \sum_{i=1}^{R} \sum_{j=1}^{C} \sum_{k=1}^{L} (|I_{i,j,k} - I'_{i,j,k}|)^2$$

Where $I_{i,j}$ (gray scale) and $I_{i,j,k}$ (color) are original image pixels and $I'_{i,j}$ and $I'_{i,j,k}$ are stego image pixels.

The comparisons of average PSNR and the average modification rate of our method with different steganographic algorithms are shown in Table II and in figure 3.

Table 2 : Comparisons of different Steganographic Algorithms using average PSNR

| Embedding Rate | Steganographic Algorithms | PSNR | Avg. rate of modification |
|---|---|---|---|
| 10% | LSBM | 61.1 | 0.05 |
| | LSBMR | **62.2** | 0.0375 |
| | **Proposed** | 56.9 | **0.0125** |
| 30% | LSBM | 56.4 | 0.15 |
| | LSBMR | **57.4** | 0.1115 |
| | **Proposed** | 54.6 | **0.0375** |
| 50% | LSBM | 54.2 | 0.25 |
| | LSBMR | **55.2** | 0.1875 |
| | **Proposed** | 53.1 | **0.0625** |
| 80% | LSBM | 53.0 | 0.4 |
| | LSBMR | **53.1** | 0.3 |
| | **Proposed** | 52.6 | **0.1** |
| 100% | LSBM | **52.2** | 0.5. |
| | LSBMR | **52.2** | 0.375 |
| | **Proposed** | 52.1 | **0.125** |

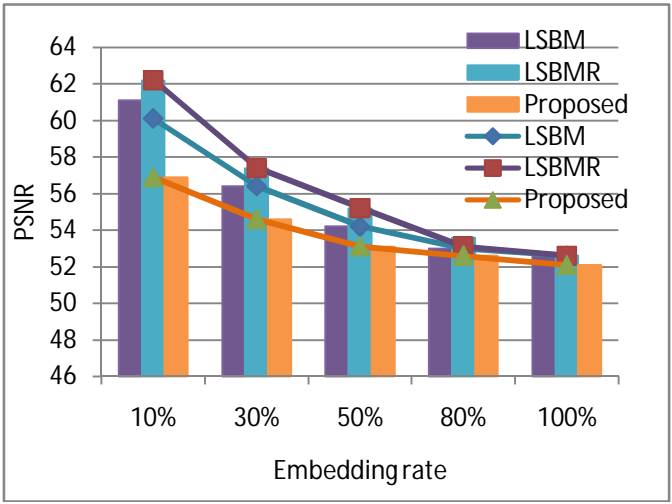

Figure 3 : Comparisons of PSNR of our methods with existing steganographic methods

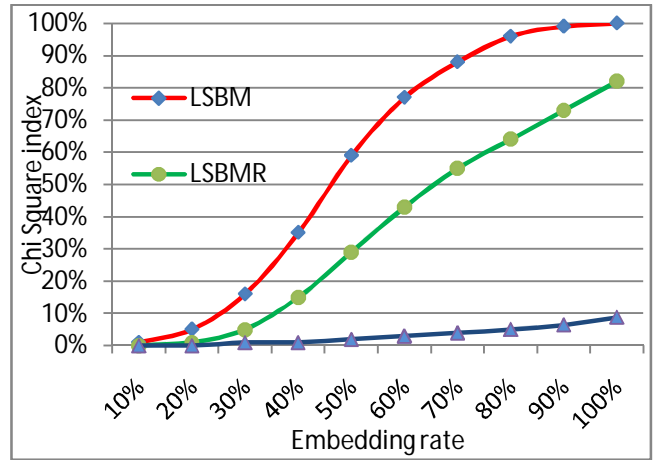

Figure 4 : Experimental result of Chi-Square attack on LSBM, LSBMR and our proposed method

The maximum gray level that can be altered in our method at maximum embedding rate is 18. So it is expected that PSNR of our algorithm will decrease comparing other algorithms. But the PSNR is still over 50 and the distortion can't be spotted in naked eyes. If we look the trend towards increasing embedding rate, PSNR our method is almost same as the best PSNR method. It indicates that our algorithm works better when embedding rate is high. The average rate of modification is less than all the existing algorithms at the same embedding rate which ensures the avoidance of asymmetry artifacts introduced in other algorithms.

c) *Steganolysis Attack*

To verify the undetectability of the proposed algorithm, two statistical steganolysis attacks are used, the Chi-Square attack and Chen steganolysis. These two algorithms generally perform very well in detecting hidden message embedded in spatial domain. Pseudo random pattern is generated before embedding message in every image in image datasets. Exactly same message is embedded using LSB, LSBMR and our proposed algorithm. For Chi-Square attack varying embedding rate from 10% to 100% is also tested to measure the risk of exposure of the hidden message. Figure 4 shows the experimental results for Chi-Square attack. The x-axis shows the embedding rate from 10% to 100%. The y-axis shows the Chi-Square index which estimates the probability of hidden message existing in the image.

In figure 4, using LSB and LSBMR method, the average Chi-square index is higher than 0.6 when the embedding rate is 80%. Using the proposed method the average Chi-square index value is always below 0.08 no matter what the embedding rate is chosen.

Figure 5 and figure 6 shows the experimental results for Chen steganolysis attack for embedding rate 30% and 60%. The x-axis shows the False Positive Rate (FPR) and y-axis shows the True Positive Rate (TPR) [21].

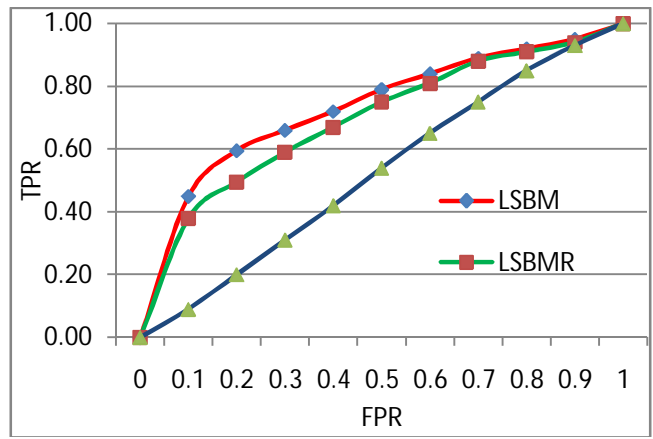

Figure 5 : Experimental results of Chen steganolysis algorithm at embedding rate 30%

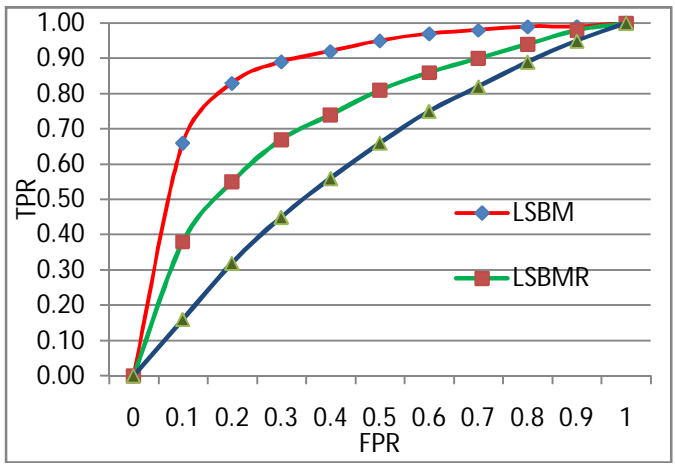

*Figure 6 :* Experimental results of Chen steganolysis algorithm at embedding rate 60%

It can be clearly observed that both steganolysis algorithms would fail in detecting embedded message in our method even when the embedding rate is high.

## V. Conclusion

In this paper existing steganographic methods in spatial domain are studied and weaknesses of existing methods are theoretically analyzed. Then this paper proposed a new data hiding method using Lucas number representation that increased the embedding capacity without degrading the quality of image. Appropriate theorems and algorithms are presented to clarify the proposed method. The experimental results are evaluated on images from well known datasets using different kinds of steganolysis algorithms. Comparisons with the existing methods show that embedding capacity, visual quality and security of stego images are improved significantly.

## VI. Acknowledgment

The authors would like to thank the anonymous referees for their constructive feedback, which helped significantly improving technical quality of this paper.

This page is intentionally left blank

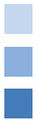